# ResNet101 and DAE for Enhance Quality and Classification Accuracy in Skin Cancer Imaging

Sibasish Dhibar
Department of Mathematics
Indian Institute of Technology Roorkee,
Roorkee, India
sdhibar@ma.iitr.ac.in

Preeti
Department of Mathematics
Indian Institute of Technology Roorkee,
Roorkee, India
preeti@ma.iitr.ac.in

**Abstract:** Skin cancer is a crucial health issue that requires timely detection for higher survival rates. Traditional computer vision techniques face challenges in addressing the advanced variability of skin lesion features, a gap partially bridged by convolutional neural networks (CNNs). To overcome the existing issues, we introduce an innovative convolutional ensemble network approach named deep autoencoder (DAE) with ResNet101. This method utilizes convolution-based deep neural networks for the detection of skin cancer. The ISIC-2018 public data taken from the source is used for experimental results, which demonstrate remarkable performance with the different in terms of performance metrics. The methods result in 96.03% of accuracy, 95.40 % of precision, 96.05% of recall, 0.9576 of F-measure, 0.98 of AUC.

## I. Introduction

According to the World Health Organization (WHO), cancer is a leading global cause of mortality, resulting in around 10 million deaths annually. Approximately 16.7% of fatalities are attributed to cancer. The number of global cancer fatalities is expected to increase by 45% from 2008 to 2030. Skin cancer (SC) is classified as the sixth most prevalent kind of cancer. SC is a harmful kind of cancer that is distinguished by the proliferation of cells in the epidermis of the skin [1]. In SC, there are seven classifications are categories which given as basal cell carcinoma (bcc), actinic keratoses (akiec), benign keratosis-like lesion (bkl), melanoma (mel), dermatofibroma (df), and melanocytic nevi (nv), vascular lesion (vasc). The procedure of diagnosis mostly includes dermoscopic analysis, clinical screening, and histopathology explorations. Dermoscopic pictures are necessary for the diagnosis of SC, but their photographic examination presents difficulties [2]. Therefore, many studies have been made in this area to enhance the precision of finding skin cancer.

This research paper focuses on using hybrid ensemble models, namely DAE and ResNet101 architecture to effectively classify skin lesions into seven classes. In addition, several preprocessing approaches, such as augmentation and data generator with pixel normalization, are used on the HAM10000 dataset (i.e., Human Against Machine) [3]. The rest of this research paper is organized as follows: Section-II encompasses the examination of relevant literature. Section-III provides a detailed explanation of the proposed mode DAE and ResNet101. Section-IV pertains to the outcomes and analysis, which is then followed by the conclusion outlined in Section-V.

## II. Related Work

Several machine learning associate 7-class SC classification designs have been emerging, which is mainly divided into 4 major groups: transfer learning, CNN, feature-clustering, and hybrid-based models. However, in the current research focuses on DAE and ResNet101 pre-trained based models. Existing literature used pre-trained convolutional neural networks (CNNs) such as AlexNet, VGG-19,

MobileNet, and ResNet50 to extract features, necessitating dimension reduction to reduce computational workload by various researchers [4], [5]. Esteva et al., [6] introduced an improved CNN model designed to expedite the training process. Younis et al., [7] optimized hyperparameters of MobileNet pre-trained model for skin lesion classification to reduce computational overhead while maintaining good accuracy. Chaturvedi et al. [8] investigated a CNN-based ensemble method for 4-class skin cancer (SC) categorization. An unaltered CNN with human adjustment of class distributions showed rather good accuracy Gosh [9]. Xin et al. [10] investigated a VIT-associate design that utilizes multiscale patch embedding and overlapping sliding spaces. Demir et al. [11] introduced two distinct deep learning (DL) algorithms, Inception-v3 and Resnet-101 for categorizing SC as benign/malignant. The Inceptionv3 model obtained a performance of 87.42%, surpassing the 84.09% attained by the Resnet-101 architecture. Keerthana et al. [12] considered the pre-trained CNNs model with a support vector machine (SVM) classifier to enhance performance indices.

### III. Proposed Methodology

The SC category dataset was from the ISIC-Archive, as described by Tschandl et al. (2018). It contains of imbalanced images representing seven types of skin cancer, with individually image having a resolution of $128 \times 128 \times 3$ pixels. To evaluate classification efficacy, DAE and ResNet-101 neural networks are combined in an architecture.

A. Data and Preprocessing

The dataset contains 10,015 training images of dermoscopy from a varied demographic, including different ethnicities, ages, and genders. Data augmentation is used on the HAM10000 dataset before training to guarantee a sufficient balance across its classes. Classes with a greater number of training photographs will exhibit bias towards achieving better accuracy, whereas classes with fewer training images will result in lower accuracy. All images are reshaped to $224 \times 224 \times 3$, followed by standard normalization for colorization.

B. Proposed Model Architecture

*Model ResNet-101*: The classification challenge utilizes the ResNet101 structure, which is short for Residual Networks and plays a significant role in computer vision tasks. The ResNet101 network uses residual connections to allow gradients to flow straight through, preventing them from becoming zero following the application of the chain rule [13]. ResNet-101 has a total of 104 convolutional layers. The structure comprises a total of 33 layers organized into 29 blocks that utilize the output of the previous block as residual connections. These residuals serve as the primary component for the summation operator at the conclusion of each block, providing the input for the subsequent blocks. The last 4 blocks use the output of the preceding block in a convolution layer with a 1×1 filter size and a stride of 1. Subsequently, a batch normalization (BN) layer is applied for normalization purposes, with its output forwarded to the summing operator located at the end of the block.

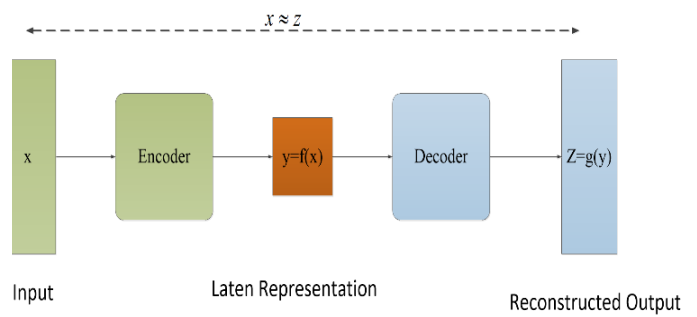

Figure 1: Architecture of DAE model

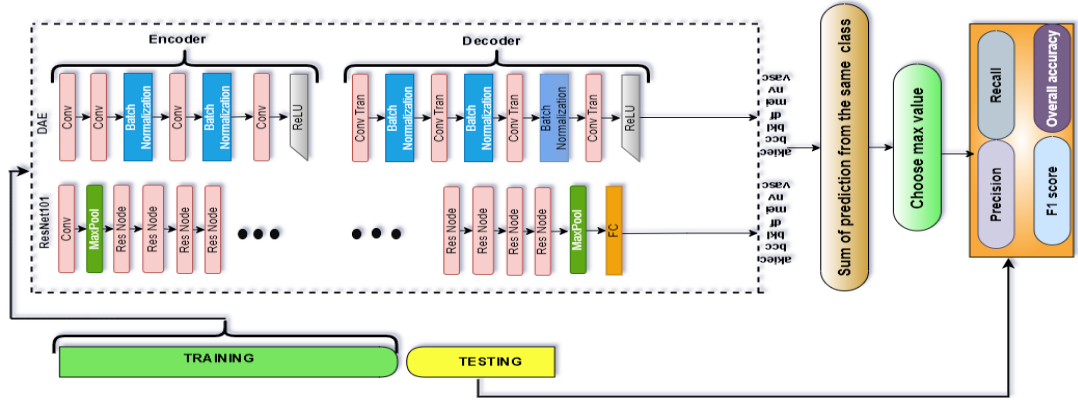

Figure 2. Proposed diagram of DAE-ResNet101-based SC classification model

*Deep autoencoder (DAE)*:

Deep neural network models can effectively organize datasets. A deep autoencoder (DAE) is an combine of ML & DL model that evaluates pictures in a dataset and replicates them inside a certain architectural framework. The loss function in the autoencoder model adjusts the model's weight parameters [14]. The architecture of DAE model is given in Figure 1. For first encode, as the value of $x$ is an input and $f(x)$ are latent space, and the second decoder, which maps the latent space onto the output $g(y)$. methods $y$ that reduces the loss rate which is set in Eq. (1). The input vector $x_n$ and an output vector $y(x_n, w)$. Therefore, the error function of the DAE model.

$$E(w) = \frac{1}{2}\sum_{n=1}^{N} \| y(x_n, w) - x_n \|^2 \quad (1)$$

*Hybrid DAE-RNT101*: The proposed model reveals the success of employing the ResNet101 pre-trained model along with deep autoencoder (DAE) models. The investigation of the proposed model consists in two main phases: initially, the dataset, which is divided into seven distinct classes, is classified directly using the ResNet101 model. This phase also includes performing the classification through the anomaly detection capabilities of the DAE method. Following this, the original dataset is reconstructed utilizing an autoencoder architecture, and the newly formed dataset is then subjected to a re-training process with the ResNet101 model. Within the framework of the proposed methodology, the dense layer, which comprises 7 features within the ResNet101 model, is utilized. These seven features undergo a classification process through an integrated approach that combines the strengths of both the DAE and ResNet101 model. The procedural flow of these initial stages is depicted in Figure 2.

### IV. Experiment and Result Discussion

We conducted a comparative analysis between the proposed DAE-ResNet101 approach and other methods, aiming to identify the highest accuracy value across various deep networks. The training process included running each approach for 100 iterations. The performance metrics of the experiment are calculated based on the given Table 1. To assess classification performance, we utilized the Python-Sklearn library's classification report utility. The Python 3.10.0 Jupyter Notebook development environment was used for all tests carried out in this research paper. A well-designed model, the accuracy should consistently improve with each epoch, while the loss per epoch should consistently decrease. The training and testing phases should exhibit similar patterns. Figure 3 presents the training and testing accuracy, as well as the training and testing loss, over 100 iterations of the proposed DAE-RNT101 model on the dataset. We have seen a minor case of overfitting in the loss curve of the proposed model (see Figure 3), especially noticeable in the elevated loss during the latter phases of training. An ideal model should have unity values for these parameters, indicating that a higher classification performance is achieved as the value approaches unity. Given that the accuracy, recall, and F1 score values in Table 2 are consistently close to 0.98, we may anticipate that the proposed models will effectively categorize the skin lesions.

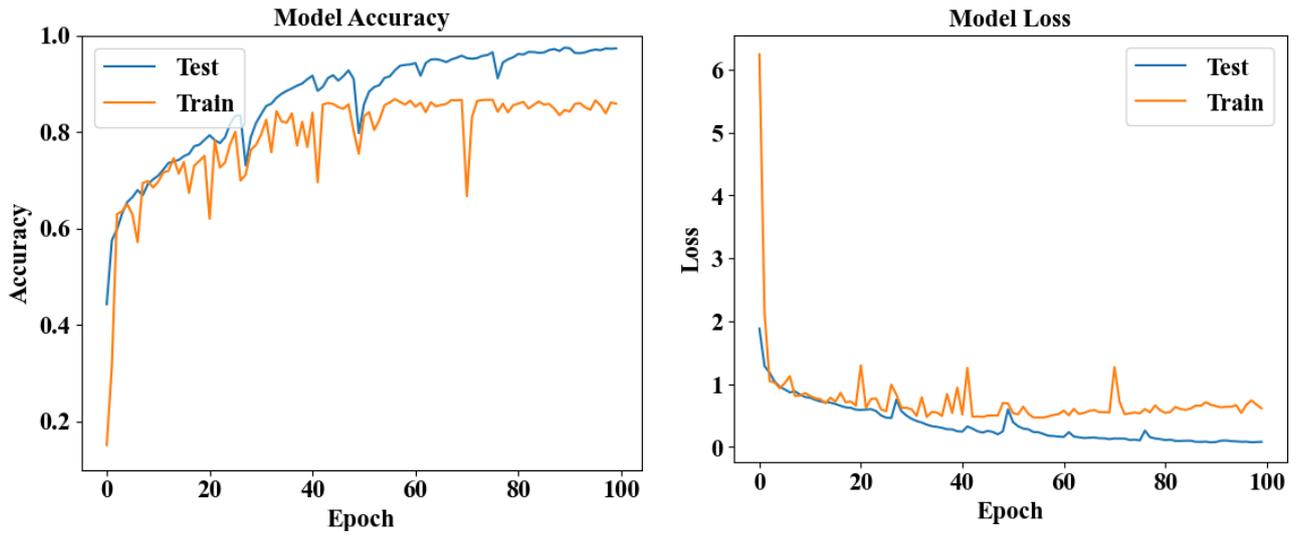

Figure 3: Accuracy and loss graphs of the for the proposed

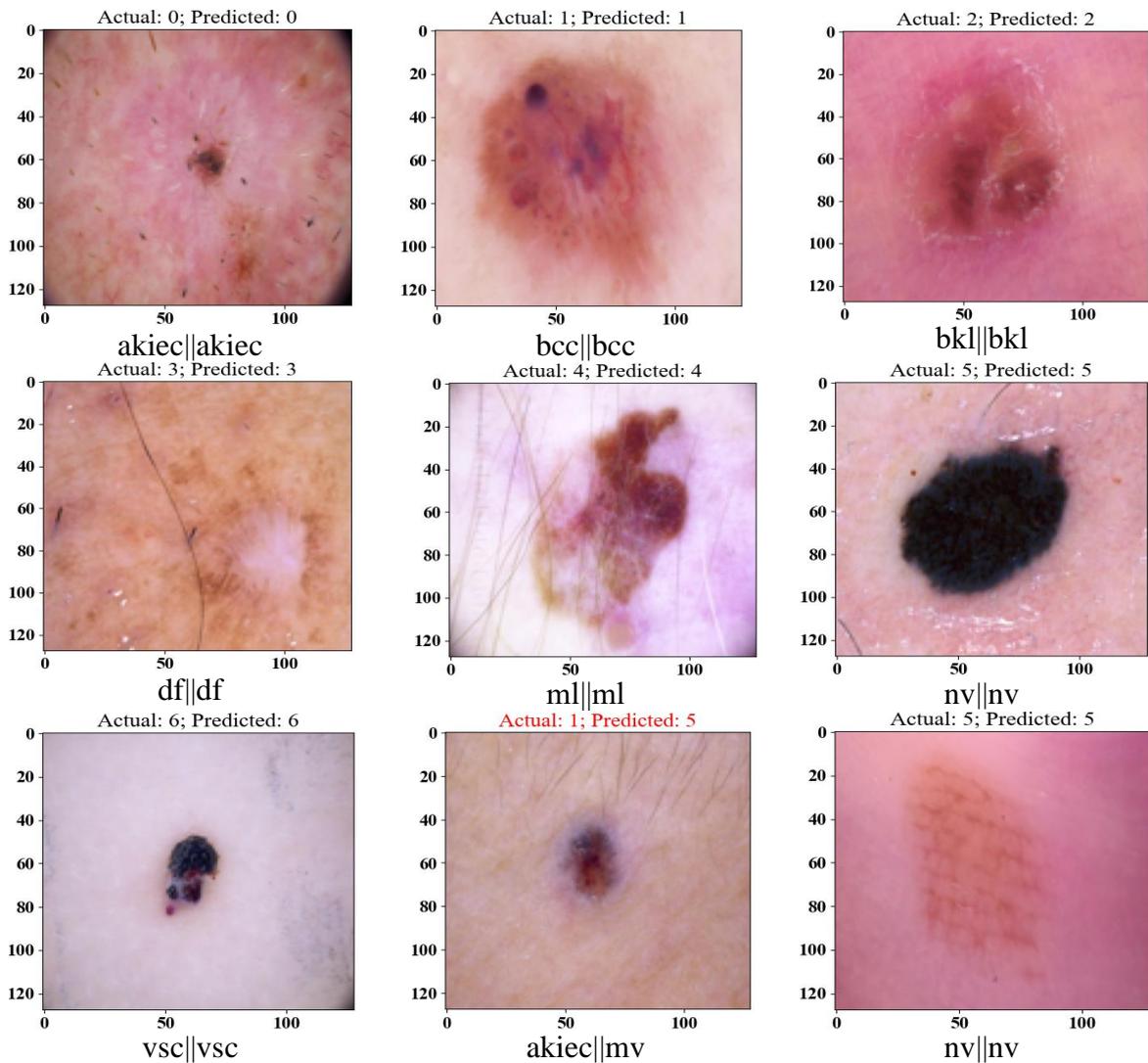

Figure 4: Illustration of 7-classified images (actual ‖predicted, the red paint contents reveal a misclassified illustration).

Table 1: Performance metrics

| Metrics | Defined as |
|---|---|
| Accuracy | $(TP+TN)/(TP+FP+TN+FN)$ |
| Recall | $TP/(TP+FN)$ |
| Precision | $TP/(TP+FP)$ |
| F1 | $(2 \times Recall \times Precision)/(Recall + Precision)$ |

Table 2: Performance of DAE-RNT101

| Accuracy: 96.03% | | | | |
|---|---|---|---|---|
| Class Label | Precision | Recall | F1 Score | AUC |
| akiec | 0.9458 | 0.9984 | 0.9213 | 0.98 |
| bcc | 0.9495 | 0.9930 | 0.9633 | 0.98 |
| bkl | 0.9468 | 0.9564 | 0.9516 | 0.96 |
| df | 0.9421 | 0.9926 | 0.9957 | 0.97 |
| mel | 0.9917 | 0.9204 | 0.9543 | 0.95 |
| nv | 0.9126 | 0.9538 | 0.9327 | 0.98 |
| vasc | 0.9897 | 0.9091 | 0.9843 | 1.00 |
| **Average** | **0.9540** | **0.9605** | **0.9576** | **0.98** |
| **STD** | **0.0280** | **0.0361** | **0.0264** | **0.015** |

Figure 4 displays the 7-class outputs of a collection of unobserved images (i.e., testing dataset). Each image has a label on top that indicates its actual class which is the image's original label as well as its predicted class. The values for each class (6=vasc, 5=nv, 4=mel, 3=df, 2=bkl, 1=bcc, 0=akiec) are shown in Figure 4. As an example, the top left picture in Figure 4 belongs to the real class of Actinic Keratoses (akiec) (Actual 0), and the ensemble architecture prediction for the same image is similarly akiec (Predicted 0). The other photos also show instances that are similar. Figure 5 illustrates the method's precision in classifying images across various categories: 46 akiec, 84 bcc, 190 bkl, 16 df, 155 mel, 887 nv, and 18 vasc images, as indicated by the diagonal entries of the confusion matrix. Moreover, the findings indicate that the classification performance of the DAE-Resnet-101 model surpasses that of the previous model, as shown in Table 3.

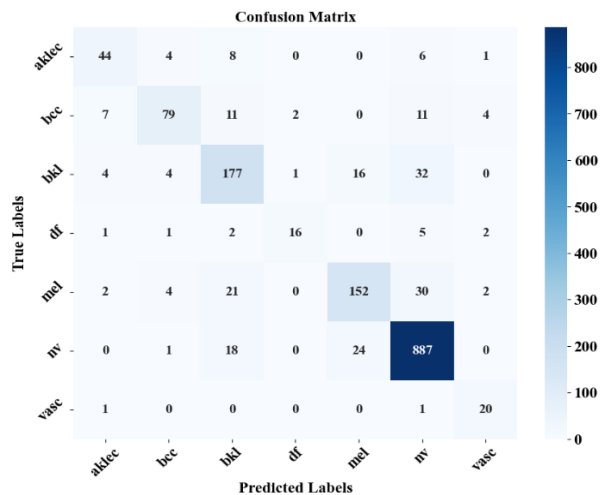

Figure 5: Confusion matrix

Table 3: Comparison with Existing Techniques

| Author | Accuracy% | F1 | Recall` |
|---|---|---|---|
| Alam et al., [15] | 91.03 | 0.8390 | 0.8336 |
| Oztel et al., [16] | 74.27 | 0.7363 | 0.7678 |
| Meswal et [17] | 92.13 | 0.93 | 0.92 |
| *DAE-ResNet101* | **96.03** | **0.9576** | **0.9605** |

## V. Conclusions and Future Direction

This research demonstrates that two distinct deep learning techniques may accurately identify skin cancer. Immediate identification and categorization of skin injuries are essential for accurate identification of skin cancer. This research work studies a novel method for detecting and categorizing skin injuries early by using combine of DAE-ResNet-101 model. The suggested DAE-ResNet-101 model achieved a average accuracy of 96.03%, with superior recall, precision, F1 score, and AUC metrics. From the existing literature, it is observed that our proposed model outperforms the existing models.